\documentclass[aps,prl,twocolumn,showpacs,preprintnumbers,superscriptaddress]{revtex4}

\usepackage{graphicx}
\usepackage{dcolumn}
\usepackage{bm}
\usepackage{amsmath}
\usepackage{amsfonts}
\usepackage{amssymb}
\usepackage[latin1]{inputenc}
\usepackage{setspace}
\usepackage[normalem]{ulem}
\usepackage{color}
\usepackage{soul}

\begin{document}

\title{Relating localized nanoparticle resonances to an associated antenna problem}

\author{Shakeeb Bin Hasan}
\address{
Institute for Condensed Matter Theory and Solid State Optics,
Friedrich-Schiller-Universit\"{a}t Jena \\ Max-Wien-Platz $1$,
$07743$, Jena, Germany}

\author{Carsten Rockstuhl}
\address{
Institute for Condensed Matter Theory and Solid State Optics,
Friedrich-Schiller-Universit\"{a}t Jena \\ Max-Wien-Platz $1$,
$07743$, Jena, Germany}

\author{Ralf Vogelgesang}
\address{Max Planck Institute for Solid State Research \\ Heisenbergstr. 1,
$70569$, Stuttgart, Germany}

\author{Falk Lederer}
\address{
Institute for Condensed Matter Theory and Solid State Optics,
Friedrich-Schiller-Universit\"{a}t Jena \\ Max-Wien-Platz $1$,
$07743$, Jena, Germany}

\begin{abstract}
We conceptually unify  the description of resonances existing at metallic
nanoparticles and optical nanowire antennas. To this end the nanoantenna is
treated as a Fabry-Perot resonator with arbitrary semi-nanoparticles forming
the terminations. We show that the frequencies of the quasi-static dipolar
resonances of these nanoparticles coincide with the frequency where the phase
of the complex reflection coefficient of the fundamental propagating plasmon
polariton mode at the wire termination amounts to $\pi$. The lowest order
Fabry-Perot resonance of the optical wire antenna occurs therefore even for a
negligible wire length. This approach can be used either to easily calculate
resonance frequencies for arbitrarily shaped nanoparticles or for tuning the
resonance of nanoantennas by varying their termination.
\end{abstract}

\pacs{41.20.Cv, 42.25.Fx, 73.22.Lp, 78.20.Bh}

\maketitle

Small particles are among the earliest cases tackled by light scattering theory.
The quasi-analytical rigorous solution for spheres dates back to the pioneering
work of Gustav Mie in $1908$ \cite{Mie1908}. This approach can be
considerably simplified if the size of the spheres is small compared to the
illuminating wavelength resulting in the quasi-static approximation
\cite{Bohren1983, Fredkin2003}. The resulting analytic formulae for the
polarizability of the sphere exhibit resonant denominators, such as the
well-known expression $\varepsilon_{\textrm{m}}(\nu) +
2\varepsilon_{\textrm{d}}(\nu) = 0$ for the dipole resonance of a metallic sphere
in a dielectric host medium. In this case and generally, the permittivities of the
spheres and their surroundings need to exhibit opposite signs at resonance. The
scattered field exhibits strongly enhanced, stationary evanescent components at
the interface -- a phenomenon which is termed localized surface plasmon
polariton (LSPP) \cite{Maier2007}. Intriguingly, it was shown that in this
approximation the quality factor of the resonance solely depends on the material
properties rather than the particle shape \cite{Wang2006} which, however,
affects the resonance frequency. Only for a few other particle shapes, e.g.
ellipsoids and spherical shells or particles with a lower dimensionality, i.e. a
cylinder, the resonance condition can be put in a similar form known from the
sphere. The exploitation of these LSPPs at nanoparticles of different shape has
led to various applications and is forming one branch of the prospering field of
plasmonics.

If the metal is a perfect conductor, another resonance is supported by metallic
wires if their length corresponds to a multiple of half the illumination
wavelength. Such metallic wires constitute the basic building blocks of
radio-frequency (RF) antennas. Recently, their downscaling into the visible
attracted considerable interest and the field of optical antennas is now similarly
established. Conceptually, optical antennas differ from RF antennas in that the
field propagating along the wire is no more purely photonic but forms another
polaritonic excitation \cite{Hopfield1958}. This type of quasiparticle is referred to
as propagating surface plasmon polariton (PSPP) due to its sole energy transport
into propagation direction. As for any guided mode phenomenon, the PSPP
dispersion relation may strongly depend on the wire's cross-section (see e.g.
\cite{Feigenbaum2006}).

The origin of resonances in these finite-length nanowires is
well-understood in terms of Fabry-Perot resonances of the PSPP mode
confined between the partially reflecting wire terminations
\cite{Novotny2007, Barnard2008, Sondergaard2008, Dorfmuller2009,
Dorfmuller2010}. Unlike in antennas at microwave frequencies, here
the reflection coefficients are complex-valued, providing an
additional phase term which mimics an increase of the wire length
and depends on the actual shape of the termination. This resembles
the situation in a planar Fabry-Perot resonator with Bragg mirrors,
where the number of layers also affects the actual phase shift and
thus the resonance condition. This is also the reason why multiples
of half the resonance wavelength differ from the wire length
\cite{Novotny2007, Bryant2008, Sondergaard2008, Dorfmuller2009,
Dorfmuller2010}. This peculiarity evoked research interest and both
analytical and numerical results on the spectral and geometrical
dependence of reflection coefficients were reported for abrupt or
flat nanowire terminations \cite{AlBader2007, Gordon2009}. Moreover,
associated geometries, such as e.g. trenches, grooves or slits on or
in flat metal surfaces or metallic thin films, were analyzed with
respect to their reflection/transmission properties of PSPP launched
along the metal surfaces \cite{Lalanne}. It allows obtaining
insights into the underlying physics of phenomena observable in such
systems. Examples thereof would be, e.g., the enhanced transmission
in subwavelength apertures \cite{LalanneNat}.

In short, to date metallic nanoparticles are largely studied in terms of LSPP
resonances whereas wire nanoantennas are commonly analyzed in terms of
PSPP standing wave phenomena and these seemingly disparate approaches
have not been systematically integrated. A few reports on variable length
nanoantennas \cite{Aizpurua2005, Bryant2008} consider spherical particles as
the limiting case of cylindrical wires. Nevertheless, it is challenging to disclose the
underlying mechanism behind this convergence of resonances in general
geometries at a physical level and to potentially exploit it for more complex
nanoparticle geometries. Here we attempt to provide a unifying view. It will turn
out that the LSPP resonances of arbitrary nanoparticles follow straightforwardly
from solving the reflection problem at the nanowire termination.

\begin{figure}[hbt]
\centering
\includegraphics[width=8.5cm]{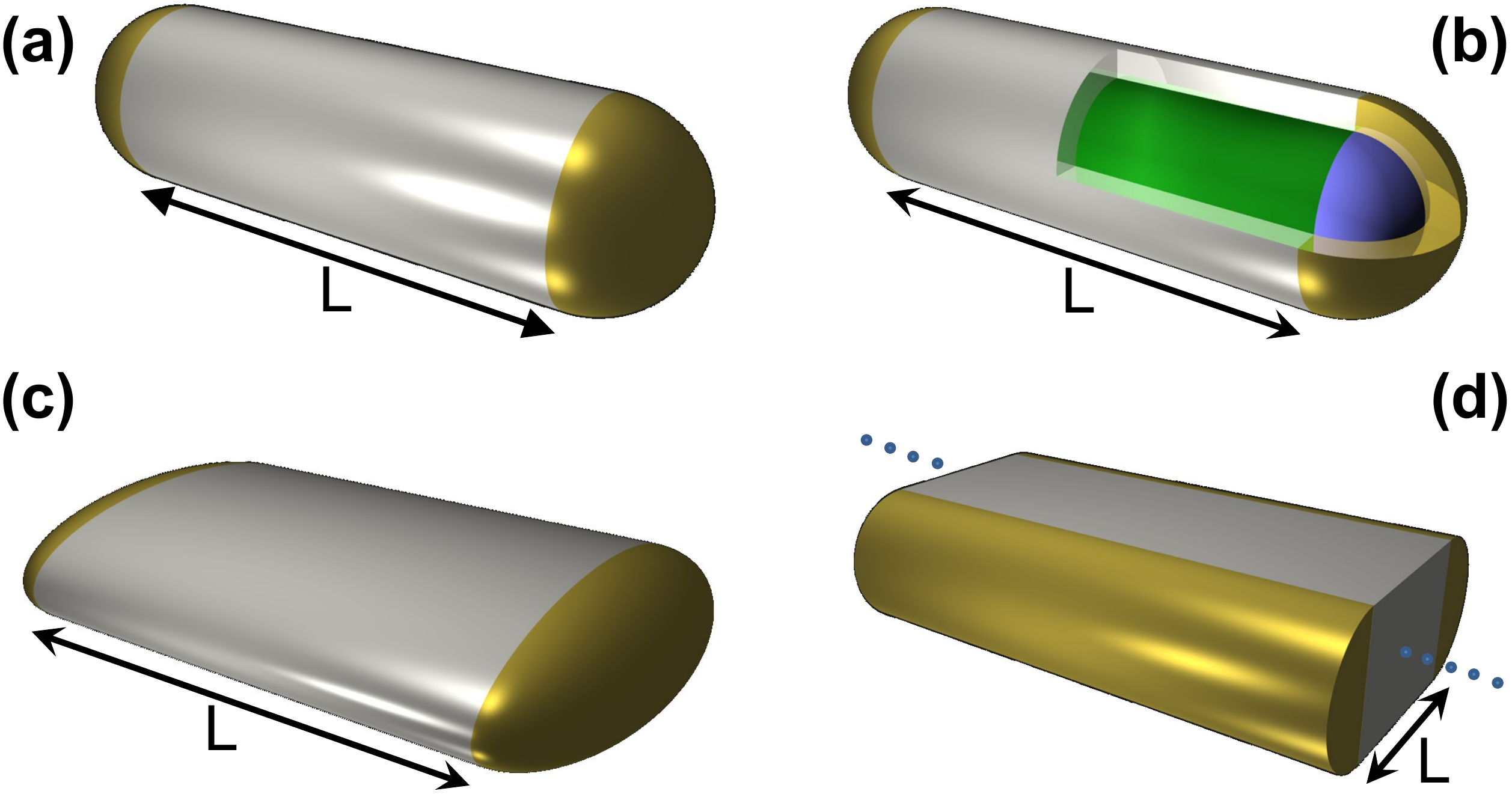}
\caption{Sketch of considered metallic nanoantenna geometries.
Circular cylinder (a) and cylindrical shell (b) with a hemispherical
termination; elliptical cylinder with a semi-elliptical termination
(c); one-dimensional nanoantenna with a semi-cylindrical
termination (d). All nanoentannas are composed of a nanowire (metallic
color) and a termination (golden) while $L$ is the nanowire's length
} \label{fig1}
\end{figure}

In the following we will consider various geometries to prove the universal
nature of our considerations, but without loss of generality we are beginning
with the circular nanowire with hemispherical terminations at both ends [Fig.
\ref{fig1}(a)]. This nanoantenna becomes a sphere for vanishing wire length $L$.
We hypothesize that the dipolar resonance of a sphere is caused by the
constructive interference of the forward and backward propagating
fundamental PSPP modes ($m=0$) of the nanowire\cite{Ashley1974,
Pfeiffer1974, Khosravi1991}. Then, the necessary condition for a Fabry-Perot
resonance -- that is, the round trip reproduction of the phase factor -- is a total
phase accumulation of an integer multiple of $2\pi$. This condition can already
be met for a wire with negligible length, i.e., the sphere, and requires a phase
shift of $\pi$ upon reflection at both terminations. In general, higher order PSPP
modes do not need to be considered since all modes with $|m|\geq2$ cut-off
below a threshold wire radius \cite{Chang2007}. Moreover, the PSPP mode with
$|m|=1$ diverges for a vanishing radius, thereby suffering from increasing
radiation losses.  Therefore,  it cannot be excited anymore \cite{Li2010}.

Thus, to generalize the idea, our claim is that it suffices to calculate the reflection
problem of a PSPP at the respective termination and to look for a $\pi$ phase
shift to identify the LSPP resonance for the respective nanoparticle shape.
Although detailed in the following only for the dipolar resonance, higher order
resonances for particles with a non-spherical shape can be understood either as
higher order Fabry-Perot resonances of the lowest order PSPP mode or lowest
order Fabry-Perot resonances of higher order PSPP modes.

We used Comsol Multiphysics to numerically solve the reflection problem at the
wire termination. The metallic nanowire is assumed to be semi-infinite and it is
surrounded by a dielectric medium with a permittivity
$\varepsilon_\mathrm{d}$. The computational cell is enclosed by perfectly
matched layers to mimic an open space. Silver (Ag) was used as the metal and its
dispersive permittivity was fully considered \cite{Johnson1972}. The cylindrical
wire had a radius of 10 nm. The exact value of the radius is not important
provided that it is much less than the wavelength.

First we calculate the dispersion relation of the fundamental wire eigenmode and
subsequently use this mode as illumination of the termination of the
semi-infinite wire. Then the total (incident and scattered) field is calculated in a
plane normal to the wire axis and located at $z = 0$. The complex reflection
coefficient of this mode at the termination is extracted by using the azimuthal
magnetic field $H_{\phi,\mathrm{tot}}(\rho,z)$ and by evaluating the overlap
integral
\begin{equation}
r = -\exp^{-i2\beta l}\frac{\int_0^\infty
H^{\star}_{\phi,0}(\rho,0)\left[H_{\phi,\mathrm{tot}}(\rho,0)-H_{\phi,0}(\rho,0)\right]\rho\,\mathrm{d}\rho}
{\int_0^\infty
H^{\star}_{\phi,0}(\rho,0)H_{\phi,0}(\rho,0)\rho\,\mathrm{d}\rho},
\end{equation}
where $H_{\phi,0}(\rho,0)$ is the $\phi$ component of the incident PSPP mode ($m=0$) at $z=0$, $\beta$
 is the associated propagation constant and  $l$ is the distance between origin ($z = 0$) and termination.
 All quantities, except the wire geometry, depend on frequency. In passing we note that the distinction of
  what belongs to the wire and what belongs to the termination is arbitrary to a certain extent. The phase
  accumulated due to propagation and the phase accumulation due to reflection can easily be merged.
  However, since we wish to discuss solely the properties of the termination, the length $l=L$ is understood
  as the length of the nanoantenna along which no change of the cross-sectional profile occurs.

\begin{figure}[hbt]
\centering
\includegraphics[width=8.5cm]{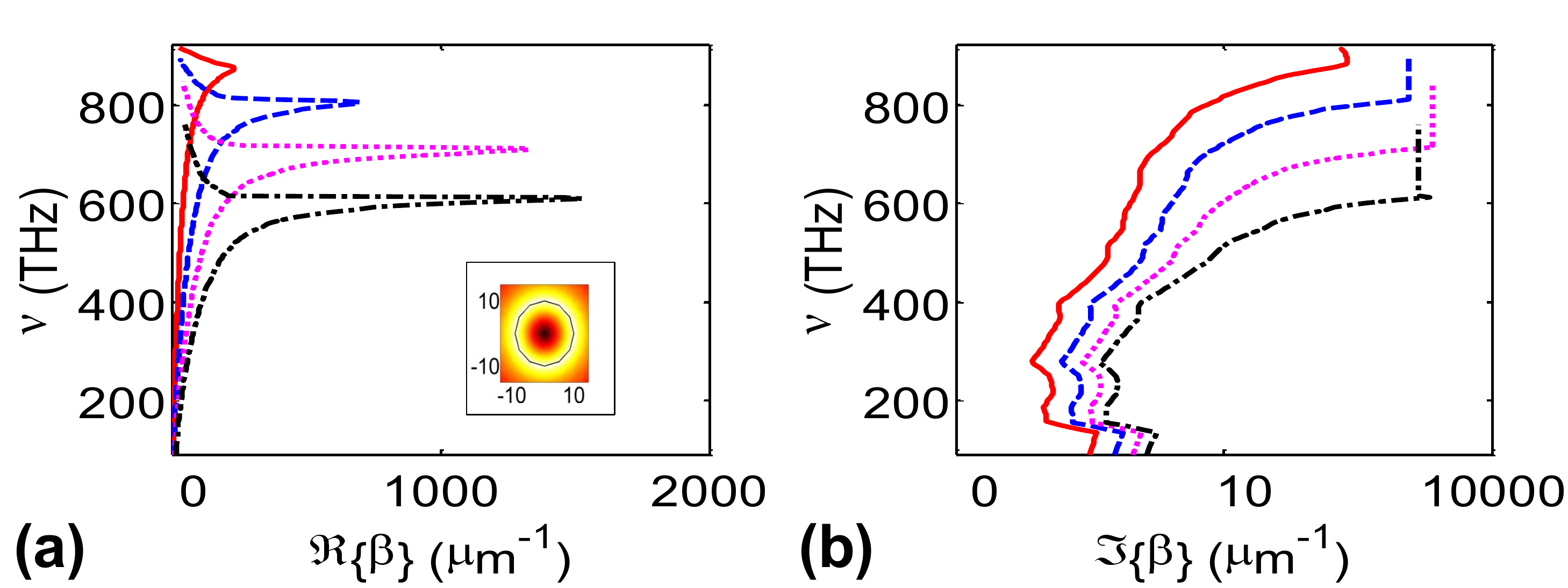}
\caption{Real (a) and imaginary part (b) of the propagation constant
$\beta$ of the lowest order PSPP mode as a function of frequency
$\nu$ for selected values of $\varepsilon_\mathrm{d}$;
$\varepsilon_\mathrm{d} = 1$ (solid red),
$\varepsilon_\mathrm{d}=2.8$ (dashed blue),
$\varepsilon_\mathrm{d}=5.8$ (dotted magenta),
$\varepsilon_\mathrm{d}=9$(dotted-dashed blue). The inset shows the
$H_{\phi}$-field norm for a core radius of $10$nm and
$\varepsilon_\mathrm{d}=1$.} \label{fig2}
\end{figure}

In Fig.~\ref{fig2} the complex-valued propagation constant $\beta$
of the lowest order PSPP mode is displayed as a function of the
frequency and the permittivity of the surrounding medium. The real
part exhibits the usual dispersion characteristic where the
propagation constant increases with frequency until back-bending
sets in. This back-bending is associated with a strongly increasing
damping (imaginary part). In the succeeding spectral domain, any
analysis of the reflection coefficient tends to be cumbersome since
dissipation will entirely dominate the system.

\begin{figure}[tbh]
\centering
\includegraphics[width=8.5cm]{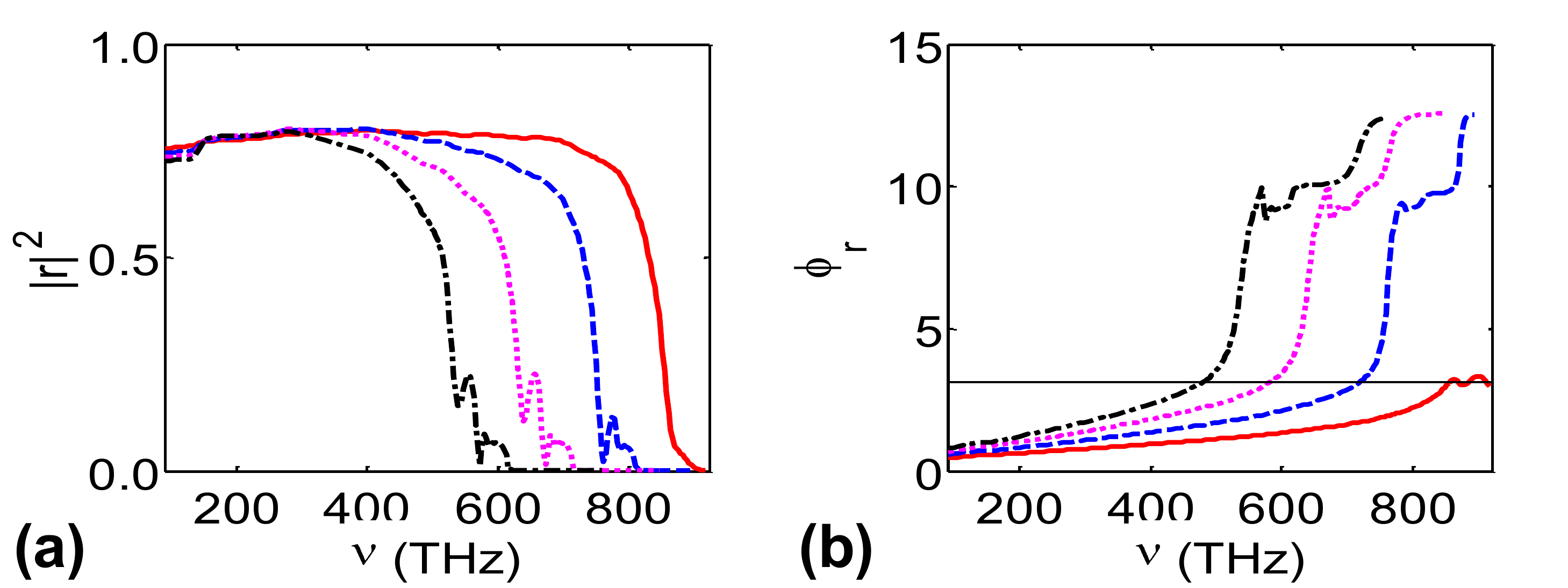}
\caption{Amplitude (a) and phase (b) of the reflection coefficient
of the nanowire for selected values of $\varepsilon_\mathrm{d}$;
$\varepsilon_\mathrm{d} = 1$ (solid red),
$\varepsilon_\mathrm{d}=2.8$ (dashed blue),
$\varepsilon_\mathrm{d}=5.8$ (dotted magenta),
$\varepsilon_\mathrm{d}=9$ (dotted-dashed blue). The horizontal
black line in (b) serves as a guide to eye and indicates where the
phase of the reflected amplitude corresponds to $\pi$.} \label{fig3}
\end{figure}

Figure \ref{fig3} shows the complex reflection coefficient as a function of
frequency for different $\varepsilon_d$ extracted from the simulation of a
semi-infinite wire for the respective lowest order PSPP mode. It can be seen that
at low frequencies the modulus is constant and large with a phase shift around
zero, suggesting a perfect metal-like behavior. The phase increases with
frequency and undergoes an abrupt change at a critical frequency. This jump is
associated with the decrease in the reflected amplitude and it appears in the
frequency interval where back bending occurs. Now it is easy to extract the
frequency where the phase jump of $\pi$ occurs and to compare it to the
resonance frequency predicted by the quasi-static theory for a small sphere.
Additionally, it can be seen that the phase surpasses even values corresponding
to multiples of $\pi$. Such frequencies would be associated to higher-order
resonances as sustained by the nanoparticle with an even higher quality factor.
Usually, such resonances are not observed, because they are dipole forbidden
and exhibit excessive damping.

\begin{figure}[hbt]
\centering
\includegraphics[width=8.5cm]{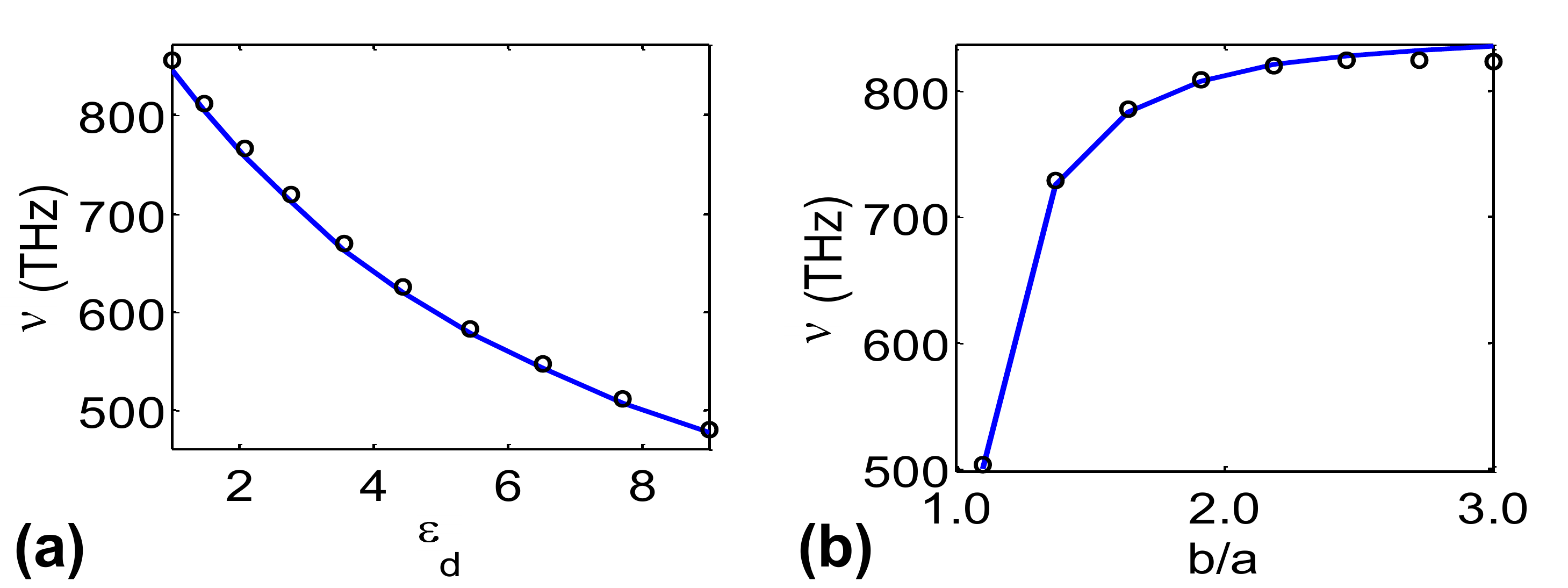}
\caption{Resonance frequency of a sphere (a) with a radius of 10nm
as a function of permittivity of surrounding medium and of a
spherical core shell particle (b) as a function of the ratio between
shell and core radii $b/a$. The core radius was kept constant at
$a=10$nm. The permittivity of both the core and the surrounding
medium was set to  $\varepsilon_\mathrm{d}=1$. Dots correspond to
resonance frequencies as extracted from the phase of the reflection
coefficients and the solid lines correspond to the predictions from
quasi-static theory.} \label{fig4}
\end{figure}

Figure \ref{fig4}(a) shows a comparison of the resonance frequencies predicted
by the quasi-static theory as well as the PSPP reflection calculation for different
surrounding media. The excellent agreement between both approaches
demonstrates that the resonances of a sphere in the quasi-static limit are directly
related to the corresponding limit of the nanoantenna problem.

To further investigate the universality of this conclusion, we
briefly analyze other particle geometries in the following. Another
special case of a spherical nanoparticle is one with a dielectric
core and a metallic shell. It is of enormous practical relevance
since its resonance frequency can be tuned across the entire visible
spectral region by varying the metallic shell thickness
\cite{Oldenburg1998}. These particles exhibit lower and higher
energy LSPP resonances which appear as a result of the hybridization
of individual resonances of the constituting metallic sphere and
dielectric void \cite{Prodan2003}. We approximate the structure
similarly as before by a nanowire made of a dielectric core
surrounded by a metallic shell terminated with a hemispherical shell
of the same construction [Fig. \ref{fig1}(b)]. Intuitively, we guess
that the lower (upper) branch of the fundamental PSPP mode of the
cylindrical metallic shell wire \cite{AlBader1993} is responsible
for the lower (higher) resonance frequency of these particles. We
repeat the aforementioned analysis for the lower branch PSPP mode
and show a similar comparison between resonance frequencies
predicted by quasi-static approximation and the reflected field
where $\phi_r=\pi$ in Fig. \ref{fig4}(b) as a function of the ratio
$b/a$ between shell and core radii, respectively. An analysis of the
highly dissipative upper branch was not attempted due to the
conceptual difficulties associated with overdamped PSPPs (see also
Fig.~\ref{fig2}).

\begin{figure}[hbt]
\centering
\includegraphics[width=8.5cm]{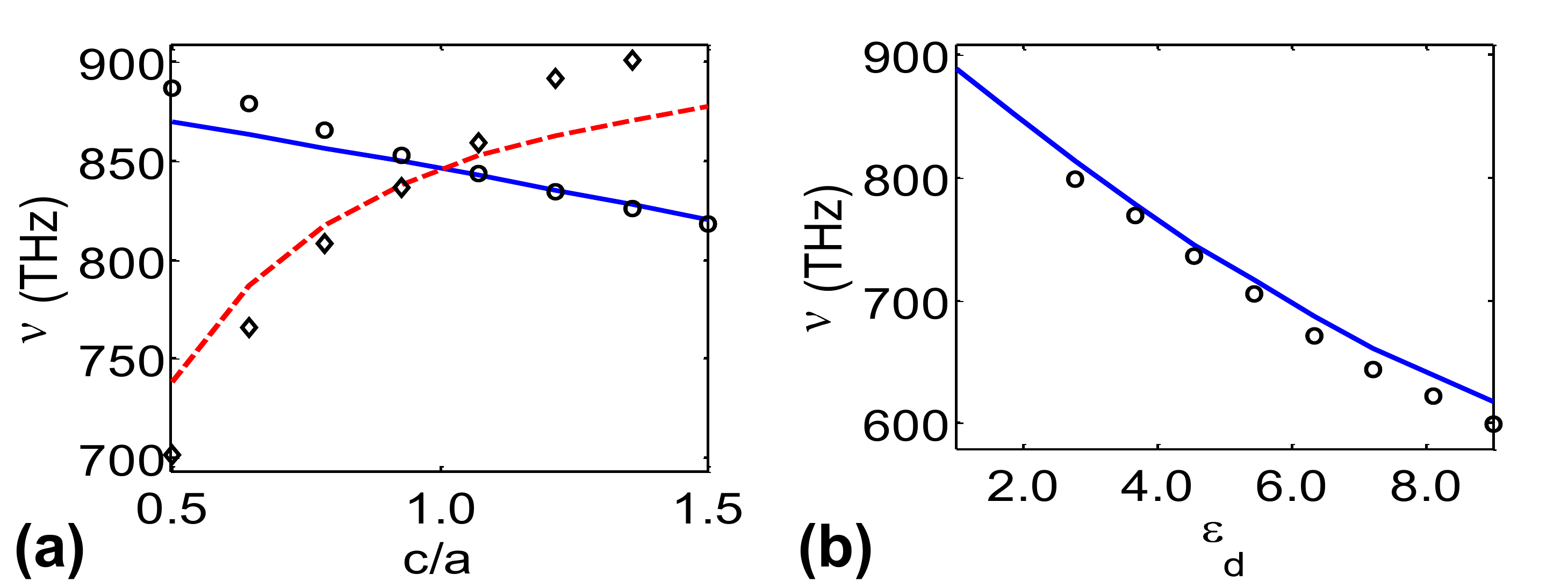}
\caption{ Resonance frequency in (a) of an ellipsoid with $b=c=20$nm
surrounded by $\varepsilon_d=1$ as a function of the ratio c/a
between semi-axes and in (b) of a cylinder with a radius of 10nm as
a function of $\varepsilon_d$. In (a), solid blue (dashed red) curve
corresponds to quasi-static resonance of the spheroid under
illuminating polarization perpendicular (parallel) to semi-axis $a$.
Circular (diamond) marks indicate frequencies at $\phi_r=\pi$ for
polarization perpendicular (parallel) to semi-axis $a$. The
polarization in the antenna simulation is selected by choosing a
respective wire mode as the illuminating field. In (b), solid blue
curve corresponds to quasi-static resonance while circular marks
denote the frequencies at which $\phi_r=\pi$. } \label{fig5}
\end{figure}

Having substantiated our principle for the case of spherical particles associated
with quasi-one-dimensional antennas, we now extend the scope of our study
toward non-spherical geometries. First we consider an elliptical geometry (see
Fig. \ref{fig1}(c)). They can be envisioned straightforwardly to be composed of
an elliptical  nanowire  (radii $a$ and $b$) of negligible length and terminations
conisting of rotational symmetric semi-ellipsoids [Fig. \ref{fig1}(c)]. Note that the
resonance frequencies of ellipsoids depend upon the illuminating
polarization\cite{Okamoto2001}. We set the semi-axis $a$ to be the symmetry
axis and present the results for the electric field polarized both perpendicular
and parallel to it. Figure \ref{fig5}(a) shows the resonance frequency as
calculated from the phase of the reflected field at the wire termination compared
to the predictions from quasi-static theory. In the antenna simulation the
polarization is chosen by selecting the eigenmode for the respective illumination.
For the  illumination perpendicular to the semi-axis $a$ (solid blue curve), we
can recognize minor deviations for small values of $b/a$. This can be attributed
to the fact that the quasi-static approximation is becoming worse for larger $a$.
However, overall we see again an excellent agreement in the predicted
resonance frequencies by both methods.

Lastly, Fig. \ref{fig5}(b) displays the comparison of the results
obtained for a one-dimensional nanoantenna with semi-cylindrical
terminations (see Fig. \ref{fig1}(d)) illuminated normal to the
cylinder axis ($k_{||}=0$) which is related to quasi-two-dimensional
antennas. As can be seen from Fig.~\ref{fig1}(d), the associated
antenna consists of an insulator-metal-insulator (IMI) waveguide
infinitely extended into one dimension. The thickness of the
nanowire is 20 nm. To form the nanoantenna, the waveguide is
terminated by semi-circular infinite cylinders [see
Fig.~\ref{fig1}(d)]. The IMI strip waveguide is well-known to
support hybridized symmetric and anti-symmetric PSPPs
\cite{Maier2007}, with the latter being strongly delocalized in the
limit of a vanishing thickness (long-range surface plasmon
polariton). The symmetric mode, on the contrary, localizes
increasingly with decreasing thickness thereby standing out as the
plausible source of LSPP resonance of cylinders in the quasi-static
limit.

Figure~\ref{fig5}(b) shows the comparison between resonance frequencies
predicted by the quasi-static limit ($\varepsilon_{\textrm{m}}(\nu) +
\varepsilon_{\textrm{d}}(\nu) = 0$) and the frequencies where $\phi_r=\pi$ for
various values of the permittivity of the surrounding dielectric. Simulations were
done by using the symmetric mode for the illumination. Again, excellent
agreement can be seen between the two approaches. We note here that we
could not observe any $\pi-$crossing in phase $\phi_r$ of reflection coefficient
for $\varepsilon_d = 1$ and $1.9$. The reason is that the resonance frequencies
occur in a spectral domain near the bulk plasma frequency where strongly
increasing absorptive losses deteriorate the analysis of a reflection coefficient.
The low propagation length of the surface plasmon polariton does not allow a
reliable analysis,which is, after all, an intrinsic problem to such plasmonic
systems. The increasing permittivity of the surrounding, however, causes a red
shift of the resonance where dissipation is reduced.

In conclusion, we have proposed a novel perspective to look at the LSPP
resonances of arbitrary metallic nanoparticles and relate them to resonances of
optical nanoantennas. We show that LSPP resonances appear at frequencies
where the phase jump upon reflection of PSPPs amounts to $\pi$. Numerical
studies show that the resonance frequencies coincide with those obtained for
LSPPs in the quasi-static approximation provided that this approximation is valid.
Whilst shedding some new light on the relation between optical antennas based
on nanoparticles and nanowires, our method establishes a systematic approach
to manipulate the resonance position of optical antennas by suitably modifying
the wire termination. This can be alternatively understood as a new degree of
freedom, in addition to the wire cross-section, in the design of nanoantennas,
which affects the propagation constant of PSPPs. {For example, it is straight
forward to imagine to equip, in perspective, nanoantennas with two different
wire terminations and to obtain a further degree of freedom to adjust the
resonance position of the nanoantenna. Moreover, we foresee that further
research devoted to the question how to relate the reflection phase to
measurable scattering quantities will be a fruitful direction. It can be anticipated
that this approach may both promote and unify the fields of localized plasmon
polaritons and optical nanoantennas.}

Financial support by the German Federal Ministry of Education and
Research (PhoNa), by the Thuringian State Government (MeMa) and the
German Science Foundation (SPP 1391 Ultrafast Nano-optics) is
acknowledged. We would like to thank S. Fahr for assistance in
preparing some figures.

\end{document}